\documentclass{WileyMSP-template}

\usepackage{amsmath}
\usepackage[utf8]{inputenc}
\usepackage{float}

\begin{document}

\pagestyle{fancy}
\rhead{\includegraphics[width=2.5cm]{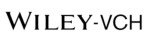}}

\title{Wedge-type engineered analog SiO$_\mathrm{x}$/Cu/SiO$_\mathrm{x}$-Memristive Devices for Neuromorphic Applications}

\maketitle

% Author: Please give full first and last names for authors and include * after the name of all corresponding authors

\author{Rouven Lamprecht*}
\author{Luca Vialetto}
\author{Tobias Gergs}
\author{Finn Zahari}
\author{Richard Marquardt}
\author{Jan Trieschmann}
\author{Hermann Kohlstedt*}

% Affiliations: Please provide adacemic titles (Prof. or Dr.) for all authors where applicable, and include an institutional email address for all corresponding authors
\begin{affiliations}
R. Lamprecht, R. Marquardt, F. Zahari, H. Kohlstedt\\
Nanoelectronics, Faculty of Engineering, Kiel University, 24143 Kiel, Germany.\\
Email Address: rola@tf.uni-kiel.de, hko@tf.uni-kiel.de\\
\medskip
H. Kohlstedt\\
Kiel Nano, Surface and Interface Science KiNSIS, Kiel University, 24118 Kiel, Germany.\\
\medskip
L. Vialetto\\
Department of Aeronautics and Astronautics, Stanford University, Stanford, CA 94305, USA\\
\medskip
 T. Gergs, J. Trieschmann\\
Theoretical Electrical Engineering, Faculty of Engineering, Kiel University, Germany.\\

\end{affiliations}

% Keywords: Please provide a minimum of three and a maximum of seven keywords, separated by commas


% Abstract should be written in the present tense and impersonal style (i.e., avoid we), and be at most 200 words long
\begin{abstract}

This study presents a comprehensive examination of the development of TiN/SiO$_\mathrm{x}$/Cu/SiO$_\mathrm{x}$/TiN memristive devices, engineered for neuromorphic applications using a wedge-type deposition technique and Monte Carlo simulations. Identifying critical parameters for the desired device characteristics can be challenging with conventional trial-and-error approaches, which often obscure the effects of varying layer compositions. By employing an \textit{off-center} thermal evaporation method, we created a thickness gradient of SiO$_\mathrm{x}$ and Cu on a 4-inch wafer, facilitating detailed resistance map analysis through semiautomatic measurements. This allows to investigate in detail the influence of layer composition and thickness on single wafers, thus keeping every other process condition constant. Combining experimental data with simulations provides a precise understanding of the layer thickness distribution and its impact on device performance. Optimizing the SiO$_\mathrm{x}$ layers to be below 12.5 nm, coupled with a discontinuous Cu layer with a nominal thickness lower than 0.6 nm, exhibits analog switching properties with an R$_\mathrm{on}$/R$_\mathrm{off}$ ratio of $>$100, suitable for neuromorphic applications, whereas R $\times$ A analysis shows no clear signs of filamentary switching. Our findings highlight the significant role of carefully choosing the SiO$_\mathrm{x}$ and Cu thickness in determining the switching behavior and provide insights that could lead to the more systematic development of high-performance analog switching components for bio-inspired computing systems.

\end{abstract}

% Text: Please use section headings and subheadings as specified below. For communications, all section headings apart from Experimental Section should be removed
% Please make the first reference to a display item bold: \textbf{Figure 1}
% Do not abbreviate Figure, Equation, etc.; display items are always singular, i.e., Figure 1 and 2.
% Equations are always singular, i.e., Equation 1 and 2, and should be inserted using the {equation} environment, not as graphics
% Please do not use footnotes in the text, additional information can be added to the Reference list.

\section{Introduction}

The persistent drive for innovation in the electronics industry has led to an exponential rise in semiconductor component integration in recent decades, significantly influenced by advancements in miniaturization and performance optimization for over fifty years \cite{M.MitchellWaldrop.2016, Strukov.2019, Moore.1965}. 
The rising power demands of contemporary Information Technology (IT), especially artificial intelligence (AI)-driven applications in smart grids and the automotive sector, challenge current CMOS technologies \cite{Shalf.2020}. Bioinspired computing paradigms emerge as a potential solution to these challenges, encompassing smart sensors to neuromorphic computing architectures, offering promising long-term prospects \cite{Mehonic.2022, DelSer.2019}. 
Memristive devices, emerging as a pivotal component in neuromorphic computing, have garnered substantial attention in recent years offering the potential to revolutionize memory storage by enabling in-memory computing as these devices are mirroring the functionality of biological synapses \cite{Ielmini.2018, Wang.2022}. These are often realised as two terminal devices in a simple metal-insulator-metal (MIM) sandwich structure which can change its resistance based on the history of the applied voltage and current \cite{Yang.2019}.
Although the precise mechanisms underlying their switching behavior are often not completely understood, memristive devices can generally be classified based on the type of mobile species within their switching layer \cite{Yang.2013}. Valence change memory (VCM) devices are mostly anion-based devices, whose resistance is changed due to the migration of corresponding anions (e.g., oxygen) \cite{Waser.2009}. Electrochemical metallization (ECM) cells consist of an electrochemically active electrode and an inert counter-electrode. The application of a voltage across them induces an electric field, which drives the migration of the mobile species, accompanying structural changes in the electrolyte. Eventually, conducting channels (filaments) form, which cause corresponding changes in resistance \cite{Waser.2009, Valov.2011, Menzel.2013}. 
Whereas these filamentary type memristive devices can fulfill the requirements of binary switching, analog-type components have the benefit of multilevel switching mimicking synaptic functions \cite{Zhang.2019}. These analog-switching memristors are gaining prominence in neuromorphic computing due to their dynamic reconfiguration capabilities, enabling the emulation of essential synaptic and neuronal functionalities. They facilitate precise weight modulation, essential for simulating key synaptic functions such as long-term potentiation, depression, and paired-pulse facilitation, enhancing the bio-realism and computational efficiency of neuromorphic systems \cite{Xi.2021, Zhang.2019b}. These devices can be utilized effectively as edge and terminal elements in artificial neural networks, playing a pivotal role in high-performance, memristor-based neuromorphic systems and the development of hardware deep neural networks \cite{Choi.2020}.
Despite the rise of alternative high-k dielectrics, such as HfO$_\mathrm{2}$, SiO$_\mathrm{2}$ remains the basic building block of CMOS-technology and thus silicon (di-)oxide and nitride-based memristors are still the subject of intensive studies in the field of their memristive properties due to their accessibility and simplified integration into existing production lines \cite{Gritsenko.2021, Mehonic.2018}. 
SiO$_\mathrm{2}$-based memristors typically operate with SiO$_\mathrm{2}$ serving merely as an electrolyte to facilitate ion migration in ECM cells \cite{Liu.2011b, Schindler.2007, Bhatt.2007, Fowler.2015, Ilyas.2022}. The utilization of copper in SiO$_\mathrm{2}$-based memristors, mainly as an active electrode material, leverages its high electrical conductivity, excellent ionization capability, and reversibility. This enables the engineering of high-performing (filamentary) devices for bipolar switching \cite{Thermadam.2010, Chen.2016}. However, coppers high diffusibility into the surrounding Si-based environment is undesirable in CMOS technology due to the risk of contamination and degradation of the semiconductor properties, making the use of copper as an electrode material unappealing. Despite this, the controlled integration of copper in SiO$_\mathrm{x}$-based memristors holds significant potential for advancing memristor technology.
Research has pivoted towards using sub-stoichiometric a-SiO$_\mathrm{x}$ as an active layer, which offers improved control over the switching properties, even without the usage of metal cation migration \cite{Yao.2010, Mehonic.2012, Chang.2012, Wang.2013, Wang.2014}. 
Engineering memristive devices for specific computational frameworks, such as neuromorphic computing, presents significant challenges. These challenges involve the meticulous selection of materials \cite{Pei.2015} and geometrical parameters like layer thicknesses and active area dimensions, which drastically affect the electric field strength and current density promoting ion diffusion and thus switching characteristics \cite{Wang.2016, Park.2015}. Despite the often limited understanding of these parameters' effects, their precise relation to device performance is crucial for reliability \cite{DiminNiu.2010, Lee.2011} and a systematic approach by wedge-like deposition has shown to be a promising method to get insight into geometrical effects \cite{Petraru.2008, Ruppelt.2014}. Systematic explorations alongside advancing device technology is essential for successful development.
Experimental efforts can be complemented by computer-aided simulations. For example, kinetic Monte Carlo (KMC) simulations may be used to determine spatially resolved film thickness for the thermal evaporation of relevant material (e.g., SiO$_\mathrm{x}$, Cu), offering precise control and insights into atomistic processes like adsorption and diffusion \cite{Zhang.2004}. Additionally KMC simulations capability in achieving uniform film thickness in multi-source thermal evaporation, highlighting its effectiveness in film uniformity control has been shown previously \cite{Chen.2010}.
The aim of this work is to establish correlations between the layers thicknesses and the switching behavior of (analog) switching memristors as well as material composition. Experimentally measured resistance maps are compared to the outcome of MC deposition simulations to determine critical layer thicknesses for obtaining analog-type switching devices.
We present a comprehensive investigation of the engineering process, characteristics and simulations of memristive device fabrication based on the insertion of copper into silicon oxide, shedding light on their potential in non-volatile memory and computing technologies as analog non-volatile memory elements in novel computing technologies such as neuromorphic computing.

\section{Methods}
\subsection{Device Fabrication}

\begin{figure}[h]
 \centering
 \includegraphics[]{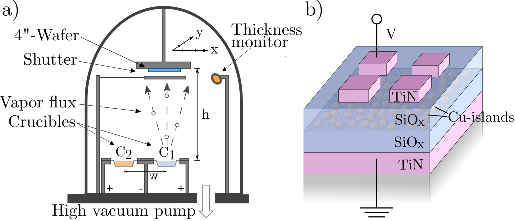}
 \caption{a) Thermal evaporation system with 2 crucible configuration for \textit{off-center} wedge-type deposition. Dimensions: w = 60 mm, h = 210 mm, crucible size: 30 mm $\times$ 10 mm. b) Device structure of the TiN/SiO$_\mathrm{x}$/Cu/SiO$_\mathrm{x}$/TiN memristive devices containing a Cu-island layer sandwiched between SiO$_\mathrm{x}$.}
 \label{fgr:ev_chamber}
\end{figure}

A specialized thermal evaporation setup was constructed to facilitate wedge-type deposition via \textit{off-center} thermal evaporation. The setup can be seen in Fig. \ref{fgr:ev_chamber} a). The first crucible (C$_\mathrm{1}$) is centered under the substrate with a substrate to crucible distance of h = 210 mm to ensure an as evenly as possible distributed deposition. The second crucible (C$_\mathrm{2}$) is located 60 mm \textit{off-center}, which leads to a non-uniform distribution of the evaporated material across the substrate, resulting in a thickness gradient of deposited films (wedge). In this work, two setups were used: Setup A: C$_\mathrm{1}$ = Cu and C$_\mathrm{2}$ = SiO, resulting in a SiO-wedge; Setup B: C$_{1}$ = SiO in and C$_{2}$ = Cu, resulting in a Cu-wedge.

The devices were fabricated on 4-inch Si wafers with 400 nm SiO$_\mathrm{2}$ (thermally oxidized) on top, patterned using optical lithography processes. The wafers are divided into 232 1 $\times$ 1 mm$^{2}$ cells across the wafer, each containing 23 subcells with six different device areas (A), ranging from 100 to 2500 $\mu$m$^{2}$, as shown in previous works \cite{Park.2021, Zahari.2019, Hansen.2015b}. The titanium nitride (TiN) electrodes were deposited in a magnetron sputtering system from a titanium (Ti)-target in a reactive process atmosphere at a power of 200 W with a gas flow of 16 sccm N$_\mathrm{2}$ and 8 sccm Ar at 1.02 $\cdot$ 10$^{-2}$ mbar process pressure and a background pressure of $\sim$ 3 $\cdot$ 10$^{-7}$ mbar. After the bottom electrode (BE) deposition, the respective wafer was immediately transferred to the evaporation system with shortest possible exposure time to the atmosphere.
The switching layer of the devices consisting of silicon oxide (SiO$_\mathrm{x}$) and copper (Cu), was then fabricated via thermal evaporation in the previously introduced dual crucible configuration. SiO$_\mathrm{x}$ is evaporated using silicon monoxide (SiO) granulate from Umicore at a base pressure of 3.2 $\cdot$ 10$^{-6}$ mbar and a deposition rate of 0.3 \r A/s (setup A) and 0.5 \r A/s (setup B), respectively. The discontinuous copper layer in between the SiO$_\mathrm{x}$ is deposited at a rate of 0.3 \r A/s for both setups. The deposition is controlled by a thickness monitor containing a gold sensor crystal of Maxtek in a Univex 300 System with an AS 053 high current source of Leybold. The top electrode (TE) was deposited via magnetron sputtering and structured via lift-off processing after transferring the wafer back to the sputter system in the shortest possible time. The device structure is sketched in Fig. \ref{fgr:ev_chamber} b). After device structuring the samples were encapsulated with 400 nm SiO by thermal evaporation. A metallization layer (100 nm Ti / 10 nm Au) was subsequently deposited to enable contacting the devices via contact pads with a semi-automatic wafer probe station.

\subsection{Electrical Measurements}

The current-voltage measurements ($I$-$V$ characteristics) were carried out with a semiautomatic measurement system to characterize the SiO$_\mathrm{x}$/Cu/SiO$_\mathrm{x}$ devices on 4"-wafers. The measurements were made with an HP 4156A parameter analyzer. The voltage was applied to the top electrode while the bottom electrode was grounded. During the $I$-$V$ characteristic measurements, the voltage was swept from 0 V to a positive maximum (forward sweep), then gradually decreased to a negative maximum, and finally returned to 0 V. For comparability - if not explicitly described otherwise - the shown measurements were performed at an electrode size of $A$ = 1225 $\mu$m$^{2}$. 
Resistance wafermaps were measured by calculating the resistance value of one device per cell using Ohm's law at a defined voltage (i.e., 0.4 V) in the forward sweep to illustrate differences in the pristine resistance values. The influence of a statistical deviation on the underlying $I$-$V$ characteristics was mitigated by considering mean values, thereby accounting for and reducing the impact of variability across individual cells. No compliance current or electro-forming step was needed before or during the measurements.

\subsection{Material Analysis}

The thickness was measured by atomic force microscopy (AFM) measurements with a SPM SmartSPM‐1000 from AIST-NT. For this, the material was deposited on a 0.25 cm$^{2}$ Si/SiO$_\mathrm{2}$ (400 nm) sample structured by photolithography and a lift-off step. Non-contact mode measurements were carried out to get information about the step height of the deposited structure for five positions across the wafer, 20 mm apart from each other. Since Cu layer thicknesses below 1 nm may not result in a continuous film, the investigated layers were produced with 20 times the intended deposition time. The resulting thicknesses were then scaled down proportionally to reflect the actual deposition time used for the analyzed devices. SiO$_\mathrm{x}$ thickness measurements were carried out under the same conditions as in the device fabrication.

\subsection{Simulation}

To disentangle the correlation of deposition parameters and resistive switching behavior, Monte Carlo simulations of transport and thin film growth have been performed. 
Conceptually similar to the approach by Chen et al.~\cite{Chen.2010}, MC transport simulations of evaporated SiO and Cu have been carried out. 
In this study, we utilize a reduced variant of an in-house PIC-DSMC plasma simulation based on the OpenFOAM framework, as described in detail elsewhere \cite{Scanlon.2010, Trieschmann.2015, Trieschmann.2018, Trieschmann.2021}. % citeweller. 
3D transport simulations are conducted on an unstructured mesh depicting the geometry shown in Fig.\,~\ref{fgr:ev_chamber} a). 
The transport was assumed to proceed without collisions, so that the simulations resemble a ray tracing algorithm. 
The basic scheme was set up as follows: Initially the evaporated flux $\Gamma$ is estimated from the maximum layer thickness measured over the wafer. This estimation was iteratively corrected in a few trial runs. Using the specified flux, the number of simulator particles to be inserted per time step $\Delta t=0.01\,\mu\text{s}$ was determined. A uniform distribution over the crucible with area $A=360\,\text{ mm}^2$ was assumed. 
Each simulator particle represented $w_\text{sp}$ physical particles. 
The locations where the simulator particles were inserted were randomly distributed over the crucible following a uniform distribution. 
Their speed was sampled from a Maxwell-Boltzmann energy distribution assuming a surface temperature of $T_s=1100\,\text{ K}$, whereas their angular distribution followed a cosine law ~\cite{Maissel.1971, Gudmundsson.2022}. 
When simulator particles reached the substrate, for each face of the mesh, the surface evolution is tracked through the surface coverage fractions $\theta_i$ and the accumulation of material $a_i$ on the surface was accounted for.~\cite{Tonneau.2018}. 
As an output of the reactor scale model, the accumulated fluxes of particles were obtained. 
The corresponding film thickness is approximated by $t_i = a_i m_i / \rho_i$, with mass density $\rho_i$ and atomic/molecular mass $m_i$.
All relevant physical quantities and parameters are collected in the Appendix table \ref{tab:simulation_params} \cite{WilliamM.Haynes.2016,Hohl.2003}.

\section{Results}

In this section, we compare experimental data with simulated outcomes to evaluate the accuracy of Monte-Carlo simulations in predicting layer thickness and material properties during memristive device fabrication at the wafer level. The experimental and simulated results are analyzed for two key parameters: layer thickness, resistivity. Furthermore analysis of the distinct devices on the wafer map examines the impact of varying layer thicknesses and sputter parameters on the production of devices exhibiting a wide range of characteristics.

\subsection{SiO$_\mathrm{x}$ wedge}

\begin{figure}[h]
 \centering
 \includegraphics[]{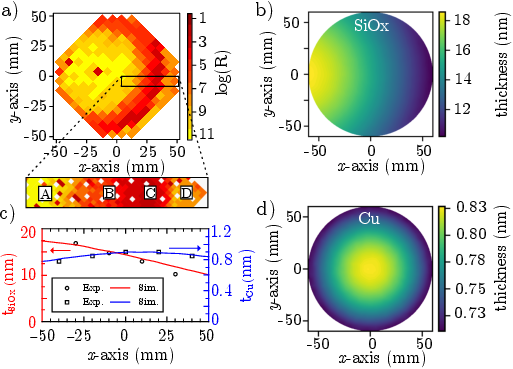}
 \caption{a) Resistance map of wedge-type deposited TiN/Si$_\mathrm{x}$/Cu/SiO$_\mathrm{x}$/TiN 4"-wafer using setup A (C$_{1}$: Cu, C$_{2}$: SiO) with SiO$_\mathrm{x}$-thickness gradient resulting in different memristive switching regions. Zoomed area with more detailed resolution and marked distinctive regions b) Monte Carlo Simulations of SiO$_{x}$ thickness/content/stoichiometry? c) Measured and simulated SiO$_\mathrm{x}$ and Cu thicknesses along the $x$-axis centerline at y = 0 mm. d) Monte Carlo Simulations of Cu thickness.}
 \label{fgr:SiOxgrad}
\end{figure}

The wedge-type (\textit{off-center}) deposition of the SiO granulate provides a SiO$_\mathrm{x}$ layer thickness (t$_{\mathrm{SiO}_x}$) ranging from 11 nm at the center to 19 nm at the edge of the wafer. When thermally evaporating from the centrally arranged copper crucible, however, there is only a nominal range of 0.7 to 0.85 nm. Both of which agree well with the outcome of Monte-Carlo simulations, which describe the statistical distribution of deposited SiO$_\mathrm{x}$ (Fig. \ref{fgr:SiOxgrad} b)) and Cu particles Fig. \ref{fgr:SiOxgrad} d)). The experimental and simulated values along the $x$-axis are shown in Fig. \ref{fgr:SiOxgrad} c). Due to the low Cu thickness, it is unlikely that a complete film is grown and island formation is assumed to dominate \cite{Zhou.2003, ZhiHuiLiu.1997, CharlesT.Campbell.1997}. %nominal values noch einmal erklären? link zu erklärung
To assess the influence of the thickness-gradient on the switching characteristics of the devices, resistance values at 0.4 V are considered -- well below the device-specific switching threshold. The results are shown in Fig.~\ref{fgr:SiOxgrad}~a). The resistance distribution resembles the SiO$_\mathrm{x}$ thickness distribution without an obvious impact of the Cu thickness (compare also the SiO$_\mathrm{x}$/Cu ration given in Fig 7, supplementary material).
The zoomed region on the $x$-axis between 0 and 50 mm reveals more detailed resistance values, encompassing all 23 devices of each cell. Along the $x$-axis, the resistance decreases from R = 10$^{11}$ $\Omega$ to R = 10$^{2}$ $\Omega$ until it rises to R = 10$^{8}$ $\Omega$ again. Between x = -50 mm and x = 0 mm the pristine resistance values are nine orders of magnitude larger. 

However, the gradual decrease and then sudden increase in resistance along the $x$-axis of the resistance map shown in \ref{fgr:SiOxgrad} a) suggests the involvement of additional factors, indicating that the observed behavior cannot be attributed solely to variations in oxide thickness. The interplay between SiO$_\mathrm{x}$ and Cu in the switching layer, along with stoichiometry changes and edge effects, may play a significant role and should be considered. However, these factors are beyond the scope of this study, which focuses on the combination of experimental and simulation finding of the impact of the layers' composition and thicknesses.

Previous studies have shown that reducing the thickness of the active layer can massively change the switching behavior of oxide based memristive devices \cite{Emelyanov.2015, Li.2017d, Mordvintsev.2018}. Ma et. al showed recently, that even in ultrathin a-SiO$_\mathrm{x}$ films (below 3 nm) memristive behavior can be observed \cite{Ma.2022}, but a systematic analysis of how varying layer thickness affects switching behavior is still missing. Even though exploring the current transport mechanism itself is out of scope of this work, the decrease in oxide resistance can be independently explained by the Poisson-Boltzmann equation (PBE), which describes the distribution of electrical potential and charge carriers within the oxide layer, thereby influencing its conductivity \cite{Abunahla.2016}. 
\begin{figure}[h]
\centering
  \includegraphics[]{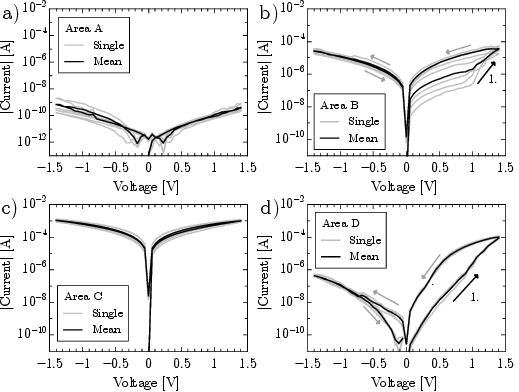}
  \caption{$I-V$ characteristics of TiN/SiO$_\mathrm{x}$/Cu/SiO$_\mathrm{x}$/TiN devices with varying thickness of SiO$_{x}$ and Cu. Single and mean curves of devices A-D in regards to the position on the wafer (see Fig. \ref{fgr:SiOxgrad}).}
  \label{fgr:4areas}
\end{figure}
Even though the thickness dependent resistance across the wafer is shown and explained before, the respective $I-V$ characteristics are considered and can be grouped into 4 different switching behaviors shown in \ref{fgr:4areas} a)-d) for area A, B, C, and D, depending on the location on the wafer (Zoomed region Fig. \ref{fgr:SiOxgrad} a)). Therefore, the mean behavior of five individual adjacent devices, to ensure device comparability regarding the layer thicknesses, are depicted in black, while the characteristics of the single devices are shown in gray to show the device-to-device variability. In the range of x = 0 mm and x = 50 mm an opening of the hysteresis associated with a resistance change can be observed and will be further analyzed later. 
% A
With a relatively high SiO$_\mathrm{x}$ thickness of t$_{\mathrm{SiO}_x}$ $>$ 13 nm and 1 nm $>$ t$_\mathrm{Cu}$ $>$ 0.7 nm (area A) the devices indicate no memristive hysteresis effects in the applied voltage region, which is reasonable due to the insulating properties of SiO$_{x}$ films and the electroforming voltages of typically more than 1.7 V \cite{Chang.2012b, Mehonic.2012}. 
% B
Lowering the  SiO$_\mathrm{x}$ thickness below t$_{\mathrm{SiO}_x}$ $<$ 12 nm and t$_\mathrm{Cu}$ = 0.8 nm (area B) shows an step-like abrupt increase of the current response for applied voltages above 1 V which is typical for filamentary ECM-cells \cite{Menzel.2013}. In the negative polarity the observed resistance change cannot be inverted, indicating a structural change of the morphology, which cannot be reversed and is therefore considered as a breakdown of the device.
% C
With SiO$_\mathrm{x}$ layer thickness of roughly t$_{\mathrm{SiO}_x}$ = 11 nm and an estimated copper thickness of t$_\mathrm{Cu}$ = 0.7 nm (area C) the $I-V$ curves show no resistance change, possibly due to a highly conductive pathway through the SiO$_\mathrm{x}$-matrix but needs to be investigated in the future.
%D
At the right edge of the Wafer with a SiO$_{x}$ layer thickness of t$_{\mathrm{SiO}_x}$ $<$ 10.6 nm and a copper content of t$_\mathrm{Cu}$ = 0.7 nm (area D) the characteristics establish a bipolar memristive hysteresis with very little device-to-device variability. The rectifying memristive behavior of the device is a desired feature for analog switching memristive devices \cite{Islam.2019b}. Even though a symmetrical device structure is designed, a asymmetrical $I-V$ characteristic and bipolar switching features of the device can be seen, which assumes a non-symmetrical layer structure possibly due to island formation during Cu deposition.
By using wedge-type depositions of the SiO$_\mathrm{x}$ and an approximately constant Cu content, particularly the dependency on the SiO$_\mathrm{x}$ thickness could be highlighted. A critical threshold thickness of t$_{\mathrm{SiO}_x}$ $<$ 10.6 nm t$_\mathrm{Cu}$ = 0.7 nm is found for analog switching devices.

\subsection{Cu wedge}

\begin{figure}[h]
 \centering
 \includegraphics[]{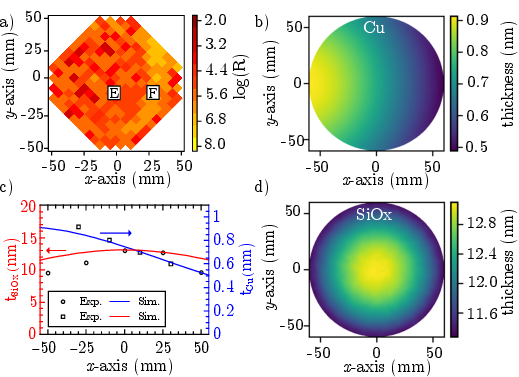}
 \caption{a) Resistance map of wedge-type deposited TiN/SiO$_\mathrm{x}$/Cu/SiO$_\mathrm{x}$/TiN 4"-wafer using setup B (C$_\mathrm{1}$: SiO, C$_\mathrm{2}$: Cu) with a Cu-thickness gradient resulting in different memristive switching regions. b) Monte Carlo Simulations of Cu thickness/content/stoichiometry? c) Measured and simulated SiO$_\mathrm{x}$ and Cu thicknesses along the $x$-axis centerline at y = 0 mm. d) Monte Carlo Simulations of SiO$_\mathrm{x}$ thickness.}
 \label{fgr:Cugrad}
\end{figure}

To get a better insight into the influence of copper regarding the memristive behavior of the devices, Fig. \ref{fgr:Cugrad} shows the resistance map with a Cu-wedge by interchanging the materials of crucible C$_\mathrm{1}$ and C$_\mathrm{2}$ during thermal evaporation (setup B, Fig. \ref{fgr:ev_chamber}). The resulting resistance wafermap is shown in Fig. \ref{fgr:Cugrad} a). Compared to the previous resistance map in Fig. \ref{fgr:SiOxgrad} a) the resistance gradient is less pronounced but still ranges from R = 10$^{2}$ $\Omega$ to R = 10$^{8}$ $\Omega$. Whereas the left half ($x$-axis -50 to 0 mm) ranges between 10$^{2}$ and 10$^{4}$, the right half ($x$-axis 0 to 50 mm) has a way more pronounced circular gradient of 10$^{4}$ to 10$^{8}$. 

Kinetic Monte Carlo Simulations in Fig. \ref{fgr:Cugrad} b) and d) visualize the SiO$_\mathrm{x}$ and Cu particle flux and calculated layer thickness on a wafer level. The simulated SiO$_{x}$-layer thickness throughout the wafer ranges from 10.4 to 12.5 nm and the estimated Cu-thickness is simulated to be from 0.5 to 0.9 nm in dependence of the wafer position. By comparison with the resistance map, the area with a higher resistance in the right end of the wafer correlates with a lower copper content.
The effect of the Cu thickness on the resistance is shown by the asymmetrical resistance distribution, which does not fit the radial distributuin of the SiO$_\mathrm{x}$-thickness depicted in Fig. \ref{fgr:Cugrad} d).
In Fig. \ref{fgr:Cugrad} c), the layer thicknesses relative to the position on the $x$-axis of the wafer for SiO$_\mathrm{x}$ and copper are illustrated. These thicknesses closely align with the experimental data obtained from separate AFM measurements conducted under identical conditions to those of the wafer under investigation (Fig. \ref{fgr:Cugrad} c)).
Since the thickness of the silicon oxide is chosen in the range of the area D in Fig. \ref{fgr:SiOxgrad} a) and in the memristive switching regime of the SiO$_\mathrm{x}$-wedge, memristive hysteresis appear all over the wafer. Due to the asymmetry in the resistance map, a copper-content-dependency in the memristive behavior can be seen in the $I-V$ characteristics, which was not as apparent for a SiO$_\mathrm{x}$-wedge. Therefore, two areas on the wafer were selected for illustration and more detailed analysis, which are labeled as area E and F in Fig. \ref{fgr:Cugrad} a). 
\begin{figure}[h]
 \centering
 \includegraphics[]{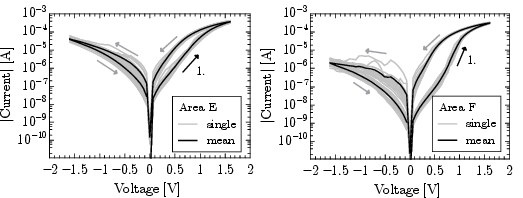}
  \caption{
  $I-V$ characteristics of TiN/SiO$_\mathrm{x}$/Cu/SiO$_\mathrm{x}$/TiN devices with varying thickness of SiO$_\mathrm{x}$ and Cu. Single (15) and mean curves of devices in area E and F in regards to the position on the wafer (see Fig. \ref{fgr:Cugrad}).}
 \label{fgr:low/highcu}
\end{figure}
For the $I-V$ characteristics shown in Fig. \ref{fgr:low/highcu}, 15 individual devices from area E are depicted in grey and the corresponding mean value is shown in black. Here the nominal thicknesses are t$_{\mathrm{SiO}_x}$ = 11.8 nm and t$_\mathrm{Cu}$ = 0.7 nm. The current in the HRS at V$_\mathrm{read}$ = 0.5 V is I$_\mathrm{HRS}$ = 1.29 $\cdot$ 10$^{-6}$ A and switches with V$_\mathrm{set}$ = 1.5 V to I$_\mathrm{LRS}$ = 1.40 $\cdot$ 10$^{-5}$ A. This corresponds to a R$_\mathrm{on}$/R$_\mathrm{off}$-ratio of 10.8. The bipolar switching behavior of the devices display a asymmetry in the hysteresis, which can be explained by the asymmetry of the device, due to the Cu-island deposition.
The devices in area F, which are located to the right end of the wafer have a nominal SiO$_\mathrm{x}$-thickness of around t$_{\mathrm{SiO}_x}$ = 11.2 nm and a Cu-thickness of t$_\mathrm{Cu}$ = 0.55 nm. The current in the HRS at Vt$_\mathrm{read}$ = 0.5 V is I$_\mathrm{HRS}$ = 9.47 $\cdot$ 10$^{-8}$ A and with V$_\mathrm{set}$ = 1.5 V rises to I$_\mathrm{LRS}$ = 1.26 $\cdot$ 10$^{-5}$ A. The following R$_\mathrm{on}$/R$_\mathrm{off}$-ratio of 133.1 is 12 times bigger than the one in area E, which leads to the assumption, that a lower Cu-content changes the resistance of the HRS of the device, since the LRS current is in the same region.
Furthermore the $I-V$ characteristics reveal an analog-like switching behavior with a gradual rise in conductance, even though there is a more intensive increase in the current in the range of 0.5 V to 1 V. No initial electro-forming or current compliance is needed to operate the devices and by applying a negative voltage of V$_\mathrm{reset}$, the device switches completely back to the HRS. 
Wedge-type deposition of the Cu-island layer shows the possibility of fine tuning the properties of the memristive device in regards of R$_\mathrm{on}$/R$_\mathrm{off}$-ratio. By lowering the Cu-content in the switching layer from t$_\mathrm{Cu}$ = 0.7 nm to t$_\mathrm{Cu}$ = 0.55 nm, the memory window can be increased by a factor of 12.

\subsection{Area dependent/analog switching}

\begin{figure}[h]
 \centering
 \includegraphics[]{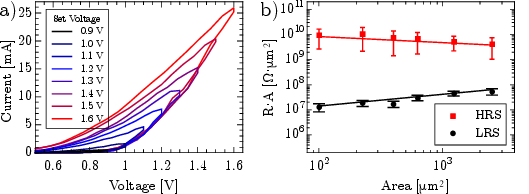}
  \caption{a) Analog set behavior between 0.9 V - 1.6 V in a linear scale, shows a gradual resistance change in dependence of the applied voltage. b) R $\times$ A plot showing area dependence of the devices. HRS (black) and LRS (red) at 0.5 V. Mean value of 5 cells (115 Devices) and standard deviation for each contact area.}
 \label{fgr:RxAanalog}
\end{figure}

By combining the results of electrical measurements and simulation, critical parameters for an analog-type memristive behavior were deduced, based on Cu-islands sandwiched between silicon oxide layers. To illustrate the analog behavior of the device, I-V curves with different maximum voltages (between 0.9 V to 1.6 V) leading to different memristive states are shown in Fig. \ref{fgr:RxAanalog} a).
Since no compliance current or forming is necessary for the switching behavior of the device and the results of the $I-V$ measurements show no abrupt change in resistance, it cannot be assumed straightaway that the switching behavior is filamentary-driven. In contrast, an homogenouse switching effect is indicated by the area-dependecy of switching shown in Fig. \ref{fgr:RxAanalog} b). 
For further analysis and understanding of the current transport mechanism, the area-dependency of the devices in area F is shown in the R $\times$ A plot in \ref{fgr:RxAanalog} b). Therefore the mean value of five cells, each containing 23 devices per feature size, were analyzed and the mean value as well as the standard deviation is depicted for a voltage of 0.4 V pre (HRS) and post (LRS) switching. 
Even if an analog switching behavior can be realized in filamentary-driven resistive random access memory (RRAM), these are limited in the number of states by stochastic processes and compatibility with Si-technology \cite{Yang.2013, Zhao.2020}. Since the number of states and $I-V$ nonlinearity are of major importance for the application in neuromorphic systems \cite{Ziegler.2018}, the reported device properties are promising for mimicking synaptic behavior.
From the graphical analysis of the $I-V$ characteristics, a filmaentary-driven process is not obvious at first glance. Since a small rise in the R $\times$ A plot is noteable in the LRS, an exclusively homogenous current transport cannot be deducted. Therefore multifilament formation \cite{Ma.2022, Kim.2023}, (Cu-)ion migration \cite{Thermadam.2010c}, interface driven processes \cite{Aoki.2014, Hansen.2015b} as well as coexistance of different processes \cite{Roy.2023}, as shown in previous works, have to be considered for future investigations of the memristive switching origin. 

\section{Conclusion}

In this work a wedge-type engineered analog TiN/SiO$_\mathrm{x}$/Cu/SiO$_\mathrm{x}$/TiN memristive device, well suited for neuromorphic applications, is presented. It was fabricated by \textit{off-center} and centered thermal evaporation of SiO$_\mathrm{x}$ and Cu, respectively, and vice versa. By fine tuning layer thicknesses, diverse resistive switching behaviors dependent on the SiO$_\mathrm{x}$ thickness were realized, ranging from analog to abrupt transition modes. By Monte Carlo Simulations a clear correlation between experimental resistance-maps and simulated layer thicknesses have been drawn. Thickness values for the shown devices have been quantified with most promising results shown for SiO$_{x}$ thicknesses below 12.5 nm. However, even minor deviations in layer thickness critically impact the switching properties of the device. This underscores the necessity for precise fabrication control and indicates that further research in this domain is essential to optimize device performance. R$_\mathrm{on}$/R$_\mathrm{off}$-ratios of 133 could be realized by combining a Cu content of less than 0.6 nm with SiO$_\mathrm{x}$ thicknesses of less than 12.5 nm. No clear indication for the predominance of filamentary or homogeneous switching has been found. Hence, further research is necessary to disentangle the potential coexistence or reveal the true origin.
The findings underscore the pivotal role of carefully choosing the material properties and parameters during deposition, particularly the SiO$_\mathrm{x}$/Cu thickness ratio. It dictates the device's switching characteristics, demonstrating that devices can be engineered to exhibit desired behaviors such as analog switching suitable for mimicking synaptic functions in neuromorphic computing systems.

% Acknowledgements
\medskip
\textbf{Acknowledgements} \par %delete if not applicable))
Funded by the Deutsche Forschungsgemeinschaft (DFG, German Research Foundation) – Project-ID 434434223 – SFB 1461
% References
\medskip

\textbf{Conflicts of interest} \par
There are no conflicts to declare.

\medskip

\textbf{Data Availability Statement} \par
The data that support the findings of this study are available from the corresponding authors upon reasonable request.

% Use the following code if you wish to generate your bibliography with BibTeX;
% replace the string "MSP-template" below with the name(s) of
% the BibTeX data base(s) you want to use.
% The resulting bibliography-output (the content of the .bbl file)
% must be pasted back into this file before submission.
% Please also include your BibTeX data base file(s) in your submission
% so that we can re-run BibTeX if necessary.
%
%\bibliographystyle{MSP}
%\bibliography{MSP-template}

% Figures/tables and captions
% Permission statements are required for all figures reproduced or adapted from previously published articles/sources. Please also ensure that all necessary permissions to reproduce images have been received
% Please remove these statements for original figures

% Please provide Biographies and photos for Essays, Feature Articles, Progress Reports, Reviews, and Perspectives for those authors who should be highlighted  
% These should be at most 100 words long
% For other article types this section can be removed
% Photographs should be 40mm broad and 50 mm high

% Table of contents entry should be 50 - 60 words long
% Image should be 55 mm broad and 50 mm high or 110 mm broad and 20 mm high

% \begin{figure}
% \textbf{Table of Contents}\\
% \medskip
%   \includegraphics{toc-image.png}
%   \medskip
%   \caption*{ToC Entry}
% \end{figure}

\footnotesize{
\bibliography{bib} %your .bib file
\bibliographystyle{MSP} %the RSC's .bst file
}
\newpage
\section{Appendix}

\begin{table*}[b]
\begin{tabular}{llllll}
\textbf{Common} &  &  &  &  &  \\
$\rho_\text{SiO}$ (g/cm$^3$) & $\rho_\text{Cu}$ (g/cm$^3$) & $m_\text{SiO}$ (amu) & $m_\text{Cu}$ (amu) &  &  \\
2.19 & 8.92 & 44.084 & 63.546 &  &  \\
\textbf{WM1} &  &  &  &  &  \\
$\Gamma_\text{SiO}$ (1/(m$^2$ s)) & $\Gamma_\text{Cu}$ (1/(m$^2$ s)) & $w_\text{sp, SiO}$ & $w_\text{sp, Cu}$ & $T_\text{SiO}$ (s) & $T_\text{Cu}$ (s) \\
5.86e20 & 1e21 & 8e9 & 6.97e8 & 180 & 27 \\
\textbf{WM2} &  &  &  &  &  \\
$\Gamma_\text{SiO}$ (1/(m$^2$ s)) & $\Gamma_\text{Cu}$ (1/(m$^2$ s)) & $w_\text{sp, SiO}$ & $w_\text{sp, Cu}$ & $T_\text{SiO}$ (s) & $T_\text{Cu}$ (s) \\
3.66e20 & 1e21 & 1e10 & 6.97e8 & 200 & 30
\end{tabular}
\caption{Parameters used for in the Monte Carlo transport simulation.}
\label{tab:simulation_params}
\end{table*}

\begin{figure}[H]
 \centering
 \includegraphics[width=8cm]{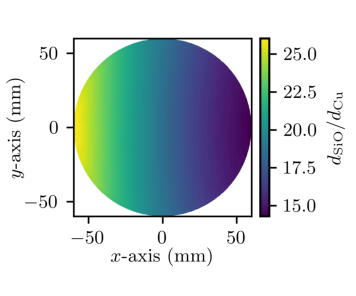}
 \caption{WM1: Film thickness ratio for SiO/Cu}
 \label{fgr:WM1_supp}
\end{figure}

\begin{figure}[H]
 \centering
 \includegraphics[width=8cm]{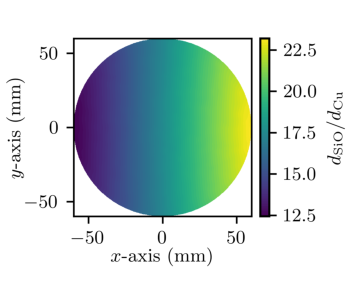}
 \caption{WM2: Film thickness ratio for SiO/Cu}
 \label{fgr:WM2_supp}
\end{figure}

\end{document}